\documentclass[12pt]{article}

\usepackage{amsmath}
\numberwithin{equation}{section}
\usepackage{mathtools}
\usepackage{dashbox}
\usepackage{a4wide}
\usepackage{epsfig}
\usepackage{theorem}
\usepackage{amssymb}
\usepackage{bbm}
\usepackage[T1]{fontenc}
\usepackage[latin1]{inputenc}
\usepackage{cancel}
\usepackage{blkarray}
\usepackage{multirow}
\usepackage{tikz}
\newcommand{\cL}{{\cal L}}

\newcommand{\bi}{\bigskip}
\newcommand{\no}{\noindent}
\newcommand{\bea}{\begin{eqnarray}}
\newcommand{\eea}{\end{eqnarray}}

\newcommand{\be}{\begin{equation}}
\newcommand{\ee}{\end{equation}}

\newcommand{\lk}{\left(}\usepackage{multirow}

\newcommand{\sli}{\sum\limits}

\newcommand{\il}{\int\limits}
\usepackage{amsmath}
\usepackage{a4wide}
\usepackage{epsfig}
\usepackage{theorem}
\usepackage{amssymb}
\usepackage{bbm}
\usepackage[T1]{fontenc}
\usepackage[latin1]{inputenc}

\begin{document}

\title{Evaluation of Some Integrals Following from $L_1$, 
the Constant of the Asymptotic Expansion of $\ln \Gamma_1 (x+1)$, 
Originating from Physics (QED)}

\author{W. Dittrich\\
Institut f\"ur Theoretische Physik\\
Universit\"at T\"ubingen\\
Auf der Morgenstelle 14\\
D-72076 T\"ubingen\\
Germany\\
electronic address: qed.dittrich@uni-tuebingen.de
}
\date{ }

\maketitle
\bi

\no

\begin{abstract}
 Comparison of three different regularization methods of calculating the one-loop effective Heisenberg-Euler Lagrangian of quantum electrodynamics 
 (QED) is employed to derive some interesting integrals involving the asymptotic expansion of $\ln \Gamma_1(x+1)$, 
 the generalized $\Gamma$ function. Here it is the constant $L_1$ that will enable us to calculate some integrals which are useful 
 in mathematics as well as in physics.
 \bi
 
 \no
The main purpose of the present article is to explicitly determine the numerical value of $L_1$ 
employing three different representations for the strong-field Heisenberg-Euler Lagrangian from which we can immediately read
off the desired value $L_1$. Given this number we can easily compute some very interesting integrals.
 \end{abstract}
 
\section{Usefulness of Riemann's Functional Equation for the Zeta Function in Physics}

Riemann introduces in his talk in Berlin in 1859 \cite[p. 147]{5}
\begin{equation}
\Gamma\left( \frac{s}{2}\right)\pi^{-\frac{s}{2}}\zeta(s) = \int_1^\infty dx \Psi(x)\left( x^{\frac{s}{2}-1} + x^{-\frac{s}{2}-\frac{1}{2}}\right) -\frac{1}{s(1-s)}\;.\label{eq:RiemannIntegralEquation}
\end{equation}
$\Psi(x)$ is related to one of Jacobi's $\theta$ functions. Notice that there is no change of the right-hand side under $s\rightarrow (1-s)$.
$\pi^{-\tfrac{s}{2}}\Gamma\left(\tfrac{s}{2}\right)\zeta(s)$ has simple poles at $s=0$ (from $\Gamma$) and $s=1$ (from $\zeta$). To remove these poles we multiply by $\tfrac{1}{2}s(s-1)$. This is the reason why Riemann defines
\begin{equation}
\xi(s) = \frac{1}{2}s(s-1)\pi^{-\frac{s}{2}}\Gamma\left(\frac{s}{2}\right) \zeta(s)\ ,\label{eq:RiemmanXi}
\end{equation}
which is an entire function ($\zeta$ is a meromorphic function). Obviously we have
\begin{equation*}
\xi(s) = \xi(1-s)
\end{equation*}
together with the symmetrical form of the functional equation which was proved by Riemann for all complex $s$:
\begin{equation}
\Gamma\left( \frac{s}{2}\right)\pi^{-\frac{s}{2}}\zeta(s) = \Gamma\left( \frac{1-s}{2}\right)\pi^{-\frac{1}{2}(1-s)}\zeta(1-s)\;.\label{eq:RiemannFunctionalEquation}
\end{equation}

Notice that the right-hand side is obtained from the left-hand side by replacing $s$ by $1-s$.

Before we continue with \eqref{eq:RiemannFunctionalEquation} we make use of two important formulae due to Euler and Legendre.

\begin{description}
\item[Legendre's duplication formula:] $\Gamma$ functions of argument $2s$ can be expressed in terms of $\Gamma$ functions of smaller arguments.
\begin{align}
\Gamma(2s) &= (2\pi)^{-\frac{1}{2}}2^{2s-\frac{1}{2}}\Gamma(s)\Gamma\left(s+\frac{1}{2}\right)\notag\\
&= \frac{1}{\sqrt{\pi}} 2^{2s-1}\Gamma(s)\Gamma\left(s+\frac{1}{2}\right)\;.\label{eq:LegendreDuplicationFormula}
\end{align}
When we replace $s\rightarrow \tfrac{s}{2}$ we obtain:
\begin{align}
\Gamma(s) &= \frac{1}{\sqrt{\pi}} 2^{s-1}\Gamma\left(\frac{s}{2}\right)\Gamma\left(\frac{s+1}{2}\right)\notag\\
\text{or}\quad \frac{\sqrt{\pi}}{2^{s-1}}\Gamma(s) &= \Gamma\left(\frac{s}{2}\right)\Gamma\left(\frac{s+1}{2}\right)\;. \label{eq:LegendreDuplicationFormulaHalfArgument}
\end{align}

\item[Euler's reflection formula for the $\Gamma$ function:] \begin{equation}
\Gamma(s)\Gamma(1-s) = \frac{\pi}{\sin \pi s}\;.\label{eq:EulersReflectionFormula}
\end{equation}
Replace $s\rightarrow\tfrac{s+1}{2}=\tfrac{s}{2}+\tfrac{1}{2}$:
\begin{align}
\Gamma\left(\frac{s+1}{2}\right)\Gamma\left(1-\frac{s+1}{2}\right) &= \frac{\pi}{\sin\left(\frac{\pi s}{2}+\frac{\pi}{2}\right)}\notag\\
\text{or}\quad \Gamma\left(\frac{s+1}{2}\right)\Gamma\left(\frac{1-s}{2}\right) &= \frac{\pi}{\cos\frac{\pi s}{2}}\quad .\label{eq:EulersReflectionFormulaShiftedArgument}
\end{align}
\end{description}

From \eqref{eq:RiemannFunctionalEquation} 
\begin{equation*}
\Gamma\left( \frac{s}{2}\right)\pi^{-\frac{s}{2}}\zeta(s) = \Gamma\left( \frac{1-s}{2}\right)\pi^{-\frac{1}{2}(1-s)}\zeta(1-s)\;.
\end{equation*}
Multiply by $\Gamma\left(\tfrac{s+1}{2}\right)$ to obtain
\begin{equation}
\pi^{-\frac{s}{2}}\Gamma\left(\frac{s}{2}\right)\Gamma\left(\frac{s+1}{2}\right)\zeta(s) = \pi^{-\frac{1-s}{2}}\Gamma\left(\frac{s+1}{2}\right)\Gamma\left(\frac{1-s}{2}\right)\zeta(1-s)\;.
\end{equation}

Then we obtain from \eqref{eq:LegendreDuplicationFormulaHalfArgument} and \eqref{eq:EulersReflectionFormulaShiftedArgument} the following equations:
\begin{align}
\pi^{-\frac{s}{2}}\frac{\sqrt{\pi}}{2^{s-1}}\Gamma(s)\zeta(s) &= \pi^{-\frac{1-s}{2}}\frac{\pi}{\cos\frac{\pi s}{2}}\zeta(1-s)\notag ,\\
\zeta(1-s) &= \frac{2}{(2\pi)^s}\cos\frac{\pi s}{2}\Gamma(s)\zeta(s)\; . \label{eq:RiemannZeta1-s}
\intertext{Here we replace $s\rightarrow 1-s$ thereby obtaining}
\zeta(s) &= \frac{2}{(2\pi)^{1-s}}\cos\frac{\pi(1-s)}{2} \Gamma(1-s)\zeta(1-s)\\
\text{or}\quad \zeta(s) &= 2^s\pi^{s-1}\sin\left(\frac{\pi s}{2}\right)\Gamma(1-s)\zeta(1-s)\; .\label{eq:RiemannZeta}
\end{align}

The latter equation will be of great importance in the following chapters.
\section{Correction of the Classical Electromagnetic Lagrangian by Vacuum Electrons}
\bi

\no
In 1936 Werner Heisenberg and Hans Euler \cite{1}
wrote down the first effective Lagrangian in quantum field theory, which incorporates a quantum correction to the classical
Lagrangian of a constant electromagnetic field; this correction is due to the polarization of the quantum vacuum (Dirac's idea), 
i.e., the effect of an external constant electromagnetic field on the motion of the vacuum electrons. To simplify matters we will 
only consider a constant magnetic field in $z$ direction. For this special case the modified Lagrangian takes the form -
in Schwinger's representation:
\begin{align}
 \label{1.1}
 \cL (B) &= \cL^{(0)} + \cL^{(1)} \, , \qquad \cL^{(0)} = - \frac{1}{2} B^2 \, , \nonumber\\
\cL^{(1)} (B) &= \frac{1}{8 \pi^2} \il^\infty_0 \frac{ds}{s^3} e^{- i  m^2 s} \left[
(e B s) \cot (e Bs) + \frac{1}{3} (eBs)^2 - 1 \right] \, .
\end{align}
 The integral was explicitly calculated for the first time in \cite{2}
 by dimensional regularization and thereafter in \cite{3}
 by the so-called zeta-function regularization. The findings of these two different methods agree exactly, whereby the result 
 obtained by the zeta-function regularization is finite without the usual subtraction of divergent counterterms. The result turns out to be
\begin{align}
 \label{1.2}
 \cL^{(1)} (B) &= - \frac{1}{32 \pi^2} \left\{ - 3 m^4 + 4 (e B)^2 \lk \frac{1}{3} - 4 \zeta' (- 1) 
 \right) + 4 m^2 (eB) (\ln 2 \pi - 1) \right. \nonumber\\
& -  2 m^4 \ln \frac{2 e B}{m^2} - 4 m^2 (eB) \ln \frac{2 e B}{m^2} - \frac{4}{3}
(eB)^2 \ln \frac{2 e B}{m^2} \nonumber\\
& \left. - 16 (eB)^2 \il^{1 + \frac{m^2}{2 eB}}_1 d x \ln \Gamma (x) \right\} \, .
\end{align}
For those values of the field strength, i.e., for strong fields $\frac{eB}{m^2} \gg 1$,
the integral over the logarithm of the gamma function only yields a constant.
\bi

\no
With  $b = \frac{eB}{m^2}$ we obtain 
\begin{align}
 \label{1.3}
 b^2 \il^{1 + \frac{1}{2 b}}_1 \ln \Gamma (x) & \approx \left. 
 b^2 \il^{1 + \frac{1}{2 b}}_1 d x \left[\ln \Gamma (1) 
 + \frac{d}{dx} \ln \Gamma (x) \right|_{x = 1} (x - 1) \right] \nonumber\\
& = b^2 \Psi (1) \il^{1 + \frac{1}{2 b}}_1 d x (x - 1) = \frac{1}{8} \Psi (1) = - \frac{1}{8} C \, .
\end{align}
Here, $C$ is Euler's number, $C = \ln \gamma = 0.577215$, and the digamma function 
\be
\label{207}
\Psi (x) = \frac{d}{dx} \ln  \Gamma (x) = \frac{\Gamma' (x)}{\Gamma (x)} \quad \mbox{at} \quad x = 1 \quad \mbox{is given by}
\quad \Psi (1) = - \ln \gamma \, . \nonumber
\ee
Therefore, by only considering the dominant forms for large magnetic field strength, we obtain for the asymptotic form of the 
one-loop effective Lagrangian in spinor QED
\begin{align}
 \label{1.4}
 \lim\limits_{\frac{e B}{m^2} \to \infty} \cL^{(1)} (B) &= - \frac{1}{32 \pi^2} \left\{ 4 (eB)^2 \lk \frac{1}{3}
 - 4 \zeta' (- 1) \right) - \frac{4}{3} (e B)^2 \ln \frac{2 eB}{m^2} \right\} \nonumber\\
 &= \frac{\alpha B^2}{6 \pi} \left\{ \ln \frac{eB}{m^2} + 12 \zeta' (- 1) - 1 + \ln 2 \right\} \, , \quad \alpha = \frac{e^2}{4 \pi} \, .
  \end{align}
On the other hand we find in Ritus' paper \cite{4} under formula number (60) the expression
\be
\label{1.5}
\lim\limits_{\frac{eB}{m^2} \to \infty} \cL^{(1)} (B) = \frac{\alpha B^2}{6 \pi} \left\{ \ln \frac{eB}{\gamma \pi m^2} 
+ \frac{6}{\pi^2} \zeta' (2) \right\} \, .
\ee
Since \eqref{1.4} and \eqref{1.5} are just two different representations of the same strong-field Lagrangian $\cL^{(1)} (B)$, 
we have the equality 
   \be
   \ln \frac{2 eB}{m^2}  - 1 + 12 \zeta' (-1) = \ln \frac{2 eB}{m^2 2 \pi \gamma}  + \frac{6}{\pi^2} \zeta' (2) \nonumber
\ee
or      
\be
\label{1.6}
-1 + 12 \zeta' (-1) = - \ln (2 \pi \gamma) + \frac{6}{\pi^2}  \zeta' (2) \, .
\ee
This important equation is in fact a direct consequence of a variant of the famous functional equation of Riemann's zeta function \cite{5}
\be
\label{1.7}
            \zeta (s) = 2^s \pi^{s-1} \sin \lk \frac{\pi s}{2} \right) \Gamma (1-s)  \zeta (1-s) \, .
\ee
              Note that the result \eqref{1.6}, which arises from long, complicated field theoretic calculations, 
              follows from solving a physics problem -
              not from analytical theory of numbers, namely by studying the behavior of vacuum electrons in presence of a constant 
              external strong classical magnetic field.
\bi
So far we considered nonlinear spinor QED where spin $\tfrac{1}{2}$ particles with mass m coupled to an external constant magnetic field.
The corresponding effective Lagrangian was given by \eqref{1.1}.

Now we want to study charged spinless particles with mass $m$, associated with a complex scalar field, which interact with a constant magnetic field.
Here the starting point is given by the Heisenberg-Euler effective Lagrangian, which in Schwinger's proper time representation reads:
\begin{equation}
\mathcal{L}_{scalar}^{(1)}(B) = -\frac{1}{16\pi^2}\int_0^\infty \frac{1}{s^3} e^{-im^2s}\left[\frac{eBs}{\sin(eBs)}-\frac{1}{6}(eBs)^2-1\right] ds\;.\label{eq:HLEffectiveLagrangianForScalarFields}
\end{equation}

Without going into detail, we just repeat the former results for spinor particles obtained for strong magnetic fields, when performing the calculation in both the $\zeta$ function
and proper time regularization. Here are the results for scalar QED:
\begin{align}
\lim_{\frac{eB}{m^2}\rightarrow\infty}\mathcal{L}_{scalar}^{(1)}(B) &= \frac{\alpha B^2}{24\pi}\left[\ln\frac{2eB}{m^2}+\left(12\zeta'(-1)-1+\ln2\right]\right]\label{eq:SFLimit1}\\
&= \frac{\alpha B^2}{24\pi}\left[\ln\frac{2eB}{m^2}+\left(-\ln\gamma\pi+\frac{6}{\pi^2}\zeta'(2)\right)\right]\label{eq:SFLimit2}
\end{align}
This brings us back to the equation \eqref{1.6}, which is equivalent to Riemann's functional equation for the zeta function. This functional equation is evidently independent of the masses involved, be they fermionic or scalar. Were it not for the factors $\tfrac{1}{24\pi}$ instead of $\tfrac{1}{6\pi}$ and $\ln\tfrac{2eB}{m^2}$ instead of $\ln\tfrac{eB}{m^2}$ in the spinor case, we could have guessed
the formulae \eqref{eq:SFLimit1} and \eqref{eq:SFLimit2}. But arriving from the proper-time integrals of \eqref{1.1} or \eqref{eq:HLEffectiveLagrangianForScalarFields} at the expressions
\eqref{1.4}, \eqref{1.5} and  \eqref{eq:SFLimit1}, \eqref{eq:SFLimit2} is a highly challenging undertaking.

Finally, let us mention that the results of \eqref{1.4} and \eqref{eq:SFLimit1} can be used in the Callan-Szymanzik renormalization equation to calculate the $\beta_\zeta(\alpha)$ function for spinor and
scalar QED to result in
\begin{equation}
\beta_\zeta(\alpha) = \frac{2}{3}\frac{\alpha}{2\pi}\;\text{(spinor)},\quad \beta_\zeta(\alpha)=\frac{1}{6}\frac{\alpha}{\pi}\;\text{(scalar)}\;.\label{eq:ResultsForScalarsAndSpinors}
\end{equation}
This confirms the correctness of the results \eqref{eq:ResultsForScalarsAndSpinors} calculated otherwise.
\no
\section{Proof of Equation \eqref{1.6} from the Functional Equation of Riemann's Zeta Function}
\bi

\no
Since equation \eqref{1.6} 
contains derivations of the $\zeta (s)$ function at $s = -1$ and $s= 2$, 
we need the derivation of \eqref{1.7}. Writing $2^s \pi^{s-1} = e^{s \ln 2} e^{(s-1) \ln \pi}$  we obtain
\begin{align}
\zeta' (s) & = (\ln 2 + \ln \pi) 2^s \pi^{s-1} \sin  \lk \frac{\pi s}{2} \right) \Gamma (1-s) \zeta (1-s) \nonumber\\
&         + \frac{\pi}{2}  2^s \pi^{s-1} \cos \lk \frac{\pi s}{2} \right) \Gamma (1-s) \zeta (1-s) \nonumber\\
&         -  2^s \pi^{s-1} \sin \lk \frac{\pi s}{2} \right) (\Gamma' (1-s) \zeta (1-s) + \Gamma (1-s) \zeta' (1-s)) \, . \nonumber
\end{align}
For $s = -2$ we find 
\begin{align}
 \label{2.1}
\zeta' (-2) & = \frac{\pi}{2}  \, 2^{-2} \pi^{-3}  \cos (-\pi) \Gamma(3)  \zeta(3) \, , \quad \Gamma (3) = 2 \, , \nonumber\\ 
\zeta' (-2) & = \frac{- \zeta (3)}{4 \pi^2} 
\end{align}                                                                                                                    
or  
\be
\zeta (3) = -4 \pi^2 \zeta' (-2) \, , \nonumber 
\ee
a result we will need later on.   
\bi

\no
For $s = -1$ we employ $\zeta (2) = \frac{\pi^2}{6}$ (Euler's Basel problem) and 
$\Gamma' (2) = 1 - C = 1- \ln \gamma, \, \Gamma (2) = 1$, so that 
\be
\zeta' (- 1) = - \ln 2 \cdot \frac{1}{12} - \ln \pi \cdot \frac{1}{12} + (1 - \ln \gamma) \cdot \frac{1}{12} +
\frac{1}{2} \cdot \frac{1}{\pi^2} \zeta'  (2) \nonumber
\ee
or
\be
\label{2.2}
\zeta' (- 1) = \frac{1}{12} \left[ 1 - \ln (2 \pi \gamma) + \frac{6}{\pi^2} \zeta' (2) \right] \, ,
\ee
which can also be written as 
\be
\label{2.3}
- 1 + 12 \zeta' (- 1) = - \ln (2 \pi \gamma) + \frac{6}{\pi^2} \zeta' (2) \, ,
\ee
which is exactly the equation that followed from the two regularization methods which produced the physically motivated results 
\eqref{1.4} and \eqref{1.5}.
\bi

\no
\section{Asymptotic Expansions of $\ln \Gamma (x+1)$ and $\ln \Gamma_1 (x+1)$}
\bi

\no
On the way to calculating the constant $L_1$ we start with the asymptotic expansion of $\ln \Gamma (x+1)$ \cite{6}:
   \begin{align}
    \label{3.1}
    \ln (x!) &= \sli^x_{x = 1} \ln x = L_0 + \lk x + \frac{1}{2} \right) \ln x - x + \frac{B_2}{1 \cdot 2 x} 
    + \frac{B_4}{3 \cdot 4 x^3} + 
    \frac{B_6}{5 \cdot 6 x^5} + \ldots \nonumber\\
    &= \ln \Gamma (x + 1) \, .
   \end{align}
This is called the Moivre-Stirling formula. Actually it was discovered by Moivre (1667-1754) with the aid of Euler's 
(1707-1783) summation formula. Stirling (1692-1770) ``only'' showed, using Wallis' (1616-1703) 
product formula, that the constant $L_0$ is given by
\be
\label{3.2}
           L_0 = \ln \sqrt{2 \pi}  = 0.918938533 \ldots \, .
                   \ee
The Bernoulli numbers in \eqref{3.1} are
\begin{align}
 \label{3.3}
B_1 & = - \frac{1}{2} , \qquad B_{2n+1}  = 0, \qquad n = 1,2,3, \ldots \, ,		\nonumber\\						  
B_2 & = \frac{1}{6} , \qquad 	B_4 = - \frac{1}{30} \, , \qquad B_6 = \frac{1}{42} \, , \qquad B_8 = - \frac{1}{30} \, \ldots \, .						
\end{align}
Useful integrals of $\ln \Gamma (x+1)$ are due to Raabe (1801-1859):
\be
\label{3.4}
\il^{x + 1}_x d t \ln \Gamma (t) = \il^x_{x - 1} dt \ln  \Gamma (t + 1) = x \ln x - x + \ln \sqrt{2 \pi} \, .
\ee
In particular,
\begin{align}
 \il^1_0 d x \ln \Gamma (x + 1) &=  \ln \sqrt{2 \pi} - 1 \, , \nonumber\\
 \il^1_0 d x \ln \Gamma (x) &= \ln \sqrt{2 \pi} \, . \nonumber
\end{align}
The generalized $\Gamma$ function $\Gamma_1 (x)$ shows up in the value of the integral
\be
\label{3.5}
\il^x_0 dt \ln \Gamma (t) = \ln \Gamma_1 (x) - \frac{x}{2} (x - 1) + x \ln \sqrt{2 \pi} \, .
\ee
One could use \eqref{3.5} as defining equation for $\ln \Gamma_1 (x)$.
Putting $x = \frac{1}{2}$ we obtain from \eqref{3.5}
\begin{align}
 \label{3.6}
 \il^{\frac{1}{2}}_0 d x \ln \Gamma (x) &= \ln \Gamma_1 (\frac{1}{2}) + \frac{1}{8} + \frac{1}{2} \ln \sqrt{2 \pi} \nonumber\\
 &= \frac{3}{2} L_1 + \frac{5}{24} \ln 2 + \frac{1}{4} \ln \pi \, ,
\end{align}
where we used 
\be
\label{3.7}
\ln \Gamma_1 (\frac{1}{2}) = \frac{3}{2} L_1 - \frac{1}{8} - \frac{1}{24} \ln 2 \, ,
\ee
which can be looked up in ref. \cite{6}. The constant $L_1$ remains to be determined, which is the main goal of this paper.
\bi

\no
Next we introduce the asymptotic expansion of $\ln \Gamma_1 (x + 1) $ \cite{6}:
\begin{align}
 \label{3.8}
 & \ln \lk 1^1 \cdot 2^2 \cdot 3^3 \ldots x^x \right) = \sli^x_{x = 1} x \ln x =: \ln \Gamma_1 (x + 1) \nonumber\\
 =& L_1 + \left[ \frac{x}{2}  (x + 1) + \frac{1}{12} \right] \ln x - \frac{1}{4} x^2 - \lk 
 \frac{B_4}{2 \cdot 3 \cdot 4 x^2} + \frac{B_6}{4 \cdot 5 \cdot 6 x^4} + \frac{B_8}{6 \cdot 7 \cdot 8 x^6} + \ldots \right) \, .
\end{align}
As pointed out in the title of this paper, the constant $L_1$ will be determined by comparing different regularization methods 
arising from physics arguments.
\bi

\no
The generalized $\Gamma$ function of the first kind satisfies the relations
\begin{align}
 \label{3.9}
 \ln \Gamma_1 (x + 1) &= \il^x_0 d t \ln \Gamma (t + 1) + \frac{x}{2} (x + 1) - x \ln \sqrt{2 \pi} \, ,\\
 \label{3.10}
 \ln \Gamma_1 (x) &= \il^x_0 d t \ln \Gamma (t) + \frac{x}{2} (x - 1) - x \ln \sqrt{2 \pi} \, .
\end{align}
$\Gamma_1 (x)$ satisfies the functional equation
\be
\label{3.11}
            \Gamma_1 (x+1) = x^x \Gamma_1 (x)
            \ee
with the constraints $\Gamma_1 (0) = \Gamma_1 (1) = \Gamma_1 (2) = 1$.
\bi

\no
The resemblance to the usual $\Gamma$ function (of the zeroth kind $\Gamma_0 \equiv \Gamma$) 
becomes obvious in observing that $\Gamma_1$
 for integer values of the argument reduces to
 \be
       \Gamma_1 (n+1) = 1^1 \cdot 2^2  \cdot 3^3 \ldots n^n\,, \qquad n > 0 \, . \nonumber
\ee
The generalized $\Gamma$ function of the $k^{th}$ kind satisfies
\begin{align}
\label{3.12}
\Gamma_{k + 1} (x + 1) &= x^{x^{k + 1}} \Gamma_{k + 1} (x) \, , \nonumber\\
\ln \Gamma_{k+1} (x+1) &= x^{k+1} \ln x + \ln \Gamma_{k+1} (x)  \, , \nonumber\\
k = 0: \quad \ln \Gamma_1 (x+1) & = x \ln x + \ln \Gamma_1 (x) \, .
\end{align}
The analogue of Raabe's formula is given by
 \be
 \label{3.13}
 \il^x_{x - 1} d t \ln \Gamma_1 (t + 1) = \frac{x^2}{2} \ln x - \frac{x^2}{4} + L_1 - \frac{1}{12} \, .
\ee
In particular,
\begin{align}
 \label{3.14}
 \il^1_0 d x \ln \Gamma_1 (x + 1) &= L_1 - \frac{1}{3} \, , \\
 \label{3.15}
 \il^1_0 dx \ln \Gamma_1 (x) &= L_1 - \frac{1}{12} \, .
\end{align}
Using $x = \frac{1}{2}$ in \eqref{3.9} and \eqref{3.7} in
\be
\label{3.16}
\ln \Gamma_1 \lk \frac{3}{2} \right) = \frac{1}{2} \ln \frac{1}{2} + \ln \Gamma_1 (\frac{1}{2})
\ee
we obtain
   \be
   \ln \Gamma_1 \lk \frac{3}{2} \right)  = - \frac{13}{24} \ln 2 + \lk \frac{3}{2} \right) L_1 - \frac{1}{8} \, , \nonumber
   \ee
so that
\begin{align}
 \label{3.17}
 \il^{\frac{1}{2}}_0 d x \ln \Gamma (x + 1) &= \ln \Gamma_1 \lk \frac{3}{2} \right) - \frac{3}{8} + \frac{1}{4} \ln 2 + \frac{1}{4} \ln \pi \nonumber\\
 &= - \frac{7}{24} \ln 2 + \frac{3}{2} L_1 - \frac{1}{2} + \frac{1}{4} \ln \pi \, .
\end{align}

\section{Numerical Value of $L_1$ from $\Gamma_1$ Function Regularization} 

Here we will make substantial use of the results contained in the rather elaborate paper of the authors of ref. \cite{7}, 
from which we can extract the $\Gamma_1$ function regularization for $\cL^{(1)} (B)$ limiting ourselves to strong fields, $eB/m^2 \gg 1$.

Here is the result: 
\be
\label{4.1}
\lim\limits_{\frac{eB}{m^2} \to \infty} \cL^{(1)} (B) = \frac{\alpha B^2}{6 \pi} \left\{ \frac{eB}{m^2} + \ln 2 - 12 L_1 \right\} \, .
\ee
Comparing the right-hand side with the results from the $\zeta$
function regularization \eqref{1.4} or from Ritus' formula \eqref{1.5}, 
the equality of the three different regularization methods teaches us
\be
\label{4.2}
           L_1 = \frac{1}{12}  - \zeta' (-1)           				
           \ee
or
              \be
              \label{4.3}
              L_1 = \frac{C}{12} + \frac{1}{12} \ln 2 \pi - \frac{1}{2 \pi^2} \zeta' (2) \, .
              \ee
Upon using   the constants
\be
   \zeta' (-1) =   -0.165421 \, , \qquad \zeta' (2) = -0.93754 \nonumber
   \ee
and Euler's constant 
\be
   \ln \gamma = C = - \psi (1) = 0.57721566490 \, , \nonumber
   \ee
we obtain
\be
\label{4.4}
           L_1 = 0.248754477 \, .
           \ee
This number can be inserted into the text wherever we meet the constant $L_1$.

Given the explicit expression \eqref{4.4}, it is worthwhile looking at the so-called Glaisher-Kinkelin constant $A$. Here is one of the many representations:
\begin{equation}
A = e^{\frac{1}{12}-\zeta'(-1)}\quad\text{from }\ln A = \frac{1}{12} - \zeta'(-1) = L_1\quad .
\end{equation}
When we take the value of $L_1$ from \eqref{4.3} we obtain:
\begin{equation}
A = (2\pi)^{\frac{1}{12}}\left[ e^{\frac{\pi^2}{6}\ln\gamma - \zeta'(2)} \right]^{\frac{1}{2\pi^2} } = 1.2824271291
\end{equation}
Here are two more representations:
\begin{equation}
\int_0^{\frac{1}{2}} \ln\Gamma(x) dx = \frac{3}{2}\ln A+\frac{5}{24}\ln 2+\frac{1}{4}\ln\pi
\end{equation}
yields 
\begin{equation}
A = 2^{-\frac{5}{36}}\pi^{-\frac{1}{6}}\exp\left[ \frac{2}{3}\int_0^{\frac{1}{2}}\ln\Gamma(x) dx\right]\quad .
\end{equation}
\begin{equation}
\int_0^{\frac{1}{2}}\ln\Gamma(x+1)dx = \frac{3}{2}\ln A -\frac{7}{24}\ln 2 + \frac{1}{4}\ln\pi -\frac{1}{2}
\end{equation}
results in
\begin{equation}
A = 2^{\frac{7}{36}} \pi^{-\frac{1}{6}} \exp\left[ \frac{1}{3}+\frac{2}{3}\int_0^{\frac{1}{2}}\ln\Gamma(x+1)dx\right] .
\end{equation}

A final remark should be made concerning the importance of $\zeta(3)$, i.e., the infinite sum of all the inverse cubes,
\begin{equation}
\zeta(3) = 1 + \frac{1}{2^3} + \frac{1}{3^3} + \frac{1}{4^3} + \frac{1}{5^3} + \ldots \, .
\end{equation}
No complex numbers are needed because $\zeta(3)$ lies in the original Euler domain of convergence, i.e., $x > 1$. No wonder that Euler became interested in the fundamentals of its value, like: Is it rational, transcendental, etc.? Remarkably it took until the late $20^{\mathrm{th}}$ century before a major breakthrough was achieved in 1979 by the 61-year-old French mathematician Roger Ap{\'e}ry, who was able to prove the irrationality of $\zeta(3)$. Today, due to his proof, the constant $\zeta(3)$ is known as Ap{\'e}ry constant. Its numerical value is given by
\begin{equation}
\zeta(3) \approx 1.202056903159\ldots \, .
\end{equation}
Hence, everywhere we meet $\zeta(3)$ in the foregoing text, one can use this value. However, $\zeta(3)$ is not only of great interest to researchers working in number theory, but is of immense value for people doing calculations in QED, like the second- and third-order (in $\alpha$) radiative corrections of the electron (and muon) anomalous magnetic moment, which is one of the best-measured and -calculated numbers in all of physics. Furthermore, $\zeta(3)$ is needed in the second-order radiative corrections to the so-called triangle anomaly, which is essential for the understanding of the $\pi^0$ decay into two photons.

Finally, while knowledge of the numerical value of Ap{\'e}ry's constant $\zeta(3) = -4 \pi^2 \zeta'(-2) \approx 1.202056903159$ is absolutely necessary for computing the just-mentioned elementary particle processes, the constants $\zeta'(-1) \approx -0.165421$ and $\zeta'(2) \approx -0.93754$  -- tied together by the relation (\ref{2.3}) -- are likewise of utmost importance for evaluating effects arising from the effective Lagrangian in QED, i.e., from the Heisenberg-Euler non-linear contribution. An example of this is the impact of the QED birefringent quantum vacuum structure on the propagation of light in the universe in the neighborhood of neutron stars, which constitute the laboratory for studying physical processes in superstrong magnetic fields.

\end{document}